\documentclass[aps,prb,twocolumn,superscriptaddress,showpacs]{revtex4-1}

\usepackage{amsmath}
\usepackage{graphicx}
\usepackage{amsfonts}
\usepackage[breaklinks=true]{hyperref}
\usepackage{hypcap}
\usepackage{verbatim}
\usepackage{color}
\usepackage{mathtools}

\def\beq{\begin{equation}}
\def\eeq{\end{equation}}

\begin{document}

\title{Temperature-dependent many-body effects in Dirac-Weyl materials: Interacting compressibility and quasiparticle velocity}
\author{F. Setiawan}
\author{S. Das Sarma}
\affiliation{Condensed Matter Theory Center and Joint Quantum
  Institute, Department of Physics, University of Maryland, College
  Park, Maryland 20742-4111, USA} 

\date{\today}

\begin{abstract}
We calculate, within the single-loop or equivalently the Hartree-Fock approximation (HFA), the finite-temperature interacting compressibility for three-dimensional (3D) Dirac materials and renormalized quasiparticle velocities for 3D and two-dimensional (2D) Dirac materials. We find that in the extrinsic (i.e., doped) system, the inverse compressibility (incompressibility) and renormalized quasiparticle velocity at $k=0$ show nonmonotonic dependences on temperature. At low temperatures the incompressibility initially decreases to a shallow minimum with a $T^2 \ln T$ dependence. As the temperature increases further, the incompressibility rises to a maximum and beyond that it decreases with increasing temperature.  On the other hand, the renormalized quasiparticle velocity at $k=0$ for both 2D and 3D Dirac materials first increases with $T^2$, rises to a maximum, and after reaching the maximum it decreases with increasing temperature. We also find that within the HFA, the leading-order temperature correction to the low-temperature renormalized extrinsic Fermi velocity for both 2D and 3D doped Dirac materials is $\ln (1/T)$.   
\end{abstract}

\pacs{71.55.Ak, 71.18.+y}
\maketitle

\section{Introduction}
Recent experimental discovery of three-dimensional (3D) Dirac semimetals~\cite{Neupane14,liu14a,xu13,Borisenko14,xu15a,liu14b, Jeon14, liang15, xu15b, xu15c, bqlv15a,bqlv15b, lxyang15,xu15d,xu15e,Sushkov15} has opened up a new avenue for realizing exotic quantum phenomena such as quantum anomalous Hall effect~\cite{gangxu11}, giant diamagnetism~\cite{Rober79,Koshino10}, oscillating quantum spin Hall effect~\cite{Wang12,Liu10}, nonlocal quantum oscillation~\cite{Potter14,Zhao15} and negative or linear quantum magnetoresistance~\cite{Abrikosov98,Zhang11,huang15,zhang15,shekhar15,He14,Feng15,Narayanan15,Novak15,Li15,xiong15}. This novel material is characterized by linear energy band dispersion in the 3D momenta near the Dirac point (i.e., the touching point between valence and conduction bands). This Dirac cone structure has been experimentally observed using ARPES measurement in several materials such as Na$_3$Bi~\cite{liu14a,xu15a,xu13}, Cd$_3$As$_2$~\cite{Borisenko14,liu14b,Neupane14,Jeon14}, TaAs~\cite{xu15b,bqlv15a,lxyang15}, NbAs~\cite{xu15c}, and TaP~\cite{xu15d,xu15e}. If the degeneracy of the Dirac cones is lifted by breaking either the time-reversal or inversion symmetry, then there are nontrivial Fermi arc surface states~\cite{Wan11,xu15b,lxyang15,xu15c,bqlv15b,bqlv15a} and unusual transport properties~\cite{Hosur13} in the resultant Weyl materials. (Since the theory developed in this paper applies to Dirac and Weyl materials with the appropriate numerical degeneracy factor correctly incorporated, in this current work we do not distinguish between them and hereafter we will refer to
both as 3D Dirac materials, systems or semimetals). In addition to transport properties, the thermodynamic property such as electronic compressibility is also of great interest as it provides information about the ground state many-body effects in interacting Dirac systems with chiral linear band dispersion. The interacting electronic compressibility has been extensively studied both theoretically and experimentally in parabolic-~\cite{Mahan,Eisenstein92,Eisenstein94,Dultz00,Ilani00,Dultz02} and linear-dispersion band materials~\cite{Hwang07,Sheehy07,Barlas07,Asgari08,Abergel09,Abergel11,Qiuzi,Martin08,Kusminskiy08,Borghi10,Yu13,Chen13,Kretinin13,Lozovik15}. In particular, the interacting compressibility in 3D metals~\cite{Mahan}, two-dimensional (2D) semiconductor structures~\cite{Eisenstein92,Eisenstein94,Dultz00,Ilani00,Dultz02}, and graphene~\cite{Martin08,Yu13,Chen13,Kretinin13} has often been the central quantity in comparing theory and experiment in the study of many-body ground state properties of strongly interacting electron systems.

In this paper, we develop a theory for the finite-temperature compressibility of a 3D Dirac material and renormalized quasiparticle velocities of 2D and 3D Dirac materials within the single-loop Hartree-Fock approximation (HFA). We also give the theoretical results for the zero-temperature Hartree-Fock (HF) and random phase approximation (RPA) compressibilities in Appendices, comparing the two approximations. We note that although we refer to our theory as the HFA (the preferred terminology in condensed matter physics), the theory is the so-called single-loop field theoretic approximation where the calculation is carried out to the leading order in the coupling constant in a perturbative sense.  The leading-order nature of the single-loop HFA also implies that the temperature-dependence we obtain would be correct to the leading order in the interaction coupling constant. We emphasize that the HFA is the leading-order perturbative expansion in the bare Coulomb interaction, which is known to produce the \textit{exact} result for the interacting compressibility to the leading order in the dimensionless coupling constant (i.e., the effective fine structure constant for the Dirac material). We further remark that for small coupling constant, the compressibilities calculated within the HFA and RPA have \textit{exactly} the same leading-order divergent term as discussed in the Appendix where we present our RPA results in terms of an expansion in the dynamically screened Coulomb interaction. 

For a 3D Dirac semimetal, the effective noninteracting low-energy Hamiltonian around a Weyl node (which for our purpose, leaving out any real spin degeneracy, is also a Dirac point where the linear conduction and valence bands meet) is 
\begin{equation}\label{eq:h0}
H_0 = v_F \mathbf{k}\cdot\boldsymbol\sigma,
\end{equation}
where $v_F$ is the noninteracting Fermi velocity (which is assumed to be momentum independent), $\mathbf{k}$ is the momentum and $\boldsymbol\sigma$ is vector of the Pauli operator. Throughout this paper, we set $\hbar = 1$. The eigenenergies of the Hamiltonian are given by $\varepsilon_s(k) = sv_Fk$ with the eigenstates (for $k_z\neq 0$)~\cite{Johannes15a} 
\begin{equation}
|\mathbf{k}s\rangle = \left(\begin{matrix}
\cos\vartheta_s/2\\
s e^{i\varphi} \sin \vartheta_s/2 
\end{matrix} \right),
\end{equation}
where $s =\pm$ denotes the conduction and valence bands, respectively, $\varphi = \arctan (k_y/k_x)$, $\cos \vartheta_+ = k_z/|\mathbf{k}|$ and $\vartheta_- \equiv \pi -\vartheta_+$. For undoped or intrinsic Dirac semimetal, the chemical potential is zero (at the Weyl node) while the doped or extrinsic Dirac semimetal has finite chemical potential. The extrinsic case is generic as the chemical potential can be easily shifted away from the Weyl node by disorder (and even for a nominally undoped system, the chemical potential is located typically either in the conduction or the valence band). Moreover, the chemical potential can also be adjusted experimentally via surface doping~\cite{liu14a}. Throughout this paper, we focus on extrinsic Dirac materials with positive Fermi energy ($E_F > 0$). At zero temperature, the noninteracting chemical potential of a 3D Dirac semimetal is the same as the Fermi energy, i.e.,
\begin{equation}
\mu_0 = E_F = v_Fk_F, 
\end{equation}
where the Fermi wave vector is given by
\begin{equation}
k_F = \left(\frac{6\pi^2 n}{g} \right)^{1/3},
\end{equation}
with $n$ and $g$ being the free carrier density and degeneracy factor (number of distinct valleys, spin, etc.), respectively. 

The compressibility is defined as the change of volume with pressure $K = -V^{-1} (\partial V /\partial P)_{T,N}$ where $K$ is the compressibility, $V$ is the volume, $P$ is the pressure, $N$ is the particle number, and $T$ is the temperature of the system. It can be shown that~\cite{Mahan,Giuliani} the inverse compressibility (incompressibility) is related to the change of chemical potential with respect to carrier density by $K^{-1} = n^2 \partial\mu/\partial n$. In 2D electron gas, the quantity $\partial\mu/\partial n$ can be obtained experimentally by measuring the quantum capacitance~\cite{Eisenstein92,Eisenstein94,Yu13,Chen13,Kretinin13} or directly measuring the density-dependent chemical potential using a scanning single-electron transistor~\cite{Ilani00,Martin08}. At zero temperature, the noninteracting incompressibility is simply
\begin{align}
\frac{1}{K_0} &= n^2 \frac{\partial\mu_0}{\partial n} = \frac{n^2}{D_0},
\end{align}
where $D_0 = \frac{1}{v_F} \left(\frac{9 g n^2}{2\pi^2} \right)^{1/3}$ is the density of states at the Fermi surface of a noninteracting 3D Dirac semimetal. For a 3D Dirac semimetal, $\partial\mu_0/\partial n \propto n^{-2/3}$ due to the linear energy dispersion while for a parabolic-band 3D electron gas (3DEG), $\partial\mu_0/\partial n \propto n^{-1/3}$.

Interaction among the electrons profoundly affects the compressibility since the chemical potential is renormalized by many-body interaction effects.  In particular, it is known that interaction could drive the electronic compressibility negative in 3D and 2D parabolic-band electron gases (as elucidated in Ref.~\cite{Mahan} and Refs.~\cite{Eisenstein92,Eisenstein94}, respectively) whereas in graphene, the prototypical 2D Dirac material, electron-electron interaction renormalizes the chemical potential, but does not lead to a negative compressibility~\cite{Hwang07}.  In the current work, we address the subject of electron-electron interaction effects on the compressibility of 3D Dirac materials.  In the usual 3D metals, the electronic temperature scale is $T_F \sim 10^4$ K (where $T_F= E_F/k_B$ is the Fermi temperature with $k_B$ being the Boltzmann constant), and hence the temperature dependence of many-body renormalization effects is not a particularly relevant topic of research in 3D metals although for lightly doped 3D semiconductors such temperature dependence may become relevant.  In 2D systems, however, both for 2D semiconductor structures and graphene, the chemical potential (i.e., the 2D carrier density) can be tuned almost arbitrarily using an external gate voltage, and therefore, temperature-dependent many-body effects take on considerable significance.  In the current work, we consider interaction effects for both the zero-temperature and finite-temperature compressibilities in 3D Dirac materials since the chemical potential can be tuned in 3D Dirac systems by extrinsic doping and for low doping, one expects the temperature dependence of the many-body renormalization effect to be significant by virtue of the Fermi energy being small.  In this work, we assume that the bare electron-electron interaction in the system is the long-range Coulomb interaction as appropriate for electrons in standard semiconductors, semimetals, and metals.

The paper is structured as follows. In Sec.~\ref{sec:finiteT}, we discuss the finite-temperature compressibility of 3D Dirac semimetals within the HFA. More specifically, we give the full numerical results of the finite-temperature compressibility and its analytic expression at low temperatures. In Sec.~\ref{sec:velocity}, we give the full numerical results and low-temperature analytic expression of the finite-temperature HF renormalized quasiparticle velocities at $k = 0$ and $k = k_F$ for 2D and 3D Dirac materials. A summary is given in Sec.~\ref{sec:summary}. In the appendices, we give detailed calculations of the zero-temperature HF compressibility of 3D Dirac materials (Appendix~\ref{sec:zeroT}) as well as the zero-temperature RPA compressibility of 2D and 3D Dirac materials (Appendix~\ref{sec:zeroRPA}). We note that the zero-temperature theoretical results are provided in the Appendices with the main text focusing on the theoretically challenging finite-temperature results.  We believe that providing both $T=0$ and finite-$T$ results in the same paper elucidate different aspects (e.g., temperature and density dependence) of the many-body renormalization effects in Dirac materials within one coherent formalism. 

\section{Finite-temperature compressibility}\label{sec:finiteT}
In this section, we present the finite-temperature compressibility of 3D Dirac materials in the single-loop (or the leading-order in interaction) theory, which is equivalent to taking into account the effect of exchange-interaction. More specifically, we focus on the extrinsic or doped Dirac semimetal with $E_F > 0$. The zero-temperature HF compressibility is given in Appendix~\ref{sec:zeroT}.

The noninteracting extrinsic chemical potential at finite temperature is determined by the conservation of total electron density:
\begin{align}\label{eq:conserv}
\frac{g}{V}\sum_{\mathbf{k}} n_F^{(+)}(\mathbf{k})+ \frac{g}{V}\sum_{\mathbf{k}}\left[n_F^{(-)}(\mathbf{k}) -1\right] = n,
\end{align}
where $n_F^{(\pm)}$ is the Fermi-Dirac distribution at temperature $T$ with energy $\varepsilon_\pm = \pm v_F k$ and $n = gk_F^3/(6\pi^2)$ is the zero-temperature carrier density due to doping. From Eq.~\eqref{eq:conserv}, we obtain the self-consistent equation for the finite-temperature noninteracting chemical potential $\mu_0$ as
\begin{equation}\label{eq:mu0}
\frac{\mu_0}{E_F}\left[(\pi t)^2 + \left(\frac{\mu_0}{E_F} \right)^2 \right] =1,
\end{equation}
where $t = T/T_F$ is the normalized temperature with $T_F = E_F/k_B$ being the Fermi temperature.
Solving Eq.~\eqref{eq:mu0}, we obtain the noninteracting chemical potential in the limit of low and high temperatures as~\cite{Johannes15b}
\begin{equation}
\frac{\mu_0}{E_F} = 
\begin{dcases}
1 - \dfrac{\pi^2 t^2}{3}, &\mbox{for } t \ll 1,\\
\dfrac{1}{\pi^2t^2}, &\mbox{for } t \gg 1. 
\end{dcases}
\end{equation}
The incompressibility is given by $1/K = n^2\partial \mu/\partial n$. Since $1/K_0 = n^2\partial E_F/\partial n$, we have $K_0/K = \partial \mu/\partial E_F = \frac{1}{v_F}\frac{\partial\mu}{\partial k_F}$. The finite-temperature noninteracting incompressibility is then given by the following analytical formulas for $T\ll T_F$ and $T\gg T_F$, respectively,
\begin{equation}\label{eq:finTcomp}
\frac{K_0}{K_{ni}(T)} = 
\begin{dcases}
 1+\dfrac{\pi^2 t^2}{3}, & \mbox{for } t \ll 1,\\
\dfrac{3}{\pi^2t^2}, &\mbox{for } t \gg 1,
\end{dcases}
\end{equation}
where $K_0$ and $K_{ni}(T)$ are the zero-temperature and finite-temperature noninteracting compressibility, respectively. Thus, to the leading-order in $t$, the incompressibility $K_{ni}(T)^{-1}$ increases quadratically and decreases inverse quadratically with temperature in the limit of low and high temperatures, respectively. This is to be contrasted with the behavior of noninteracting incompressibility of a parabolic 3DEG (i.e., normal 3D metals and doped semiconductors), where it increases quadratically and linearly with temperature in the limit of low and high temperatures, respectively~\cite{Qiuzi}. 

Within the HFA, the interacting chemical potential is the sum of kinetic energy and exchange self-energy. The exchange self-energy is given by:
\begin{equation}\label{eq:exselfenergy}
\Sigma_{s}^{\mathrm{ex}}(k) = -\sum_{s'}\int{\frac{d^3 q}{(2\pi)^3}n_F(\xi_{s'\mathbf{q}}) V_{\mathbf{k}-\mathbf{q}}F_{ss'}(\mathbf{k},\mathbf{q})},
\end{equation}
where  $s,s'=\pm$ are the band indices,  $V_{\mathbf{q}} = 4\pi e^2/(\kappa q^2)$ is the ``bare" long-range ($\sim 1/r$) Coulomb interaction in the momentum space with $\kappa$ being the background dielectric constant of the material, $\xi_{s'\mathbf{q}} = s'v_Fq-\mu_0$ is the free dispersion, $n_F(\xi_{s'\mathbf{q}})$ is the Fermi-Dirac distribution, and $F_{ss'}(\mathbf{k},\mathbf{q}) = |\langle\mathbf{k}s|\mathbf{q}s'\rangle| = (1+ ss'\cos\theta)/2$ is the overlap of the eigenstates with $\theta$ being the angle between $\mathbf{k}$ and $\mathbf{q}$. We note that the Hartree part of the self-energy vanishes exactly for the homogeneous system since it is canceled by the  positive background charge necessary for neutrality, thus leaving the exchange or the Fock self-energy as the only leading-order single-loop term. The exchange self-energy can be split into the intrinsic self-energy ($\Sigma_{s}^{\mathrm{int}}$), which arises from the contribution at zero chemical potential and the extrinsic self-energy ($\Sigma_{s}^{\mathrm{ext}}$), which takes into account the correction to the self-energy due to nonzero chemical potential:
\begin{subequations}\label{eq:sigmaintext}
\begin{align}
\Sigma_{s}^{\mathrm{int}}(k) &= - \int \frac{d^3q}{(2\pi)^3} V_{\mathbf{k}-\mathbf{q}}F_{s-}(\mathbf{k},\mathbf{q}),\\
\Sigma_{s}^{\mathrm{ext}}(k) &= - \sum_{s'}\int\frac{d^3q}{(2\pi)^3}\delta n_F(\xi_{s'\mathbf{q}})V_{\mathbf{k}-\mathbf{q}}F_{ss'}(\mathbf{k},\mathbf{q}),
\end{align}
\end{subequations}
where $\delta n_F(\xi_{s'\mathbf{q}}) = n_F(\xi_{s'\mathbf{q}})  - (1-s')/2$ is the difference between the electron occupation number of finite-temperature extrinsic 3D Dirac semimetals and the intrinsic $T= 0$ case. 

Since the chemical potential (consequently, the compressibility) is only affected by the electron energy at Fermi momentum $k_F$, in the following, we will evaluate the self-energy at $k= k_F$. Considering extrinsic 3D Dirac materials with $E_F > 0$ $(s = +1)$, we have the intrinsic self-energy as~\cite{Johannes15a}
\begin{align}
\Sigma_+^{\mathrm{int}}(k_F) =& -\frac{\alpha v_F k_F}{\pi} \left(\frac{1}{2\eta} +\frac{1}{6 \eta^2} - \frac{1}{3} \ln \left|1-\frac{1}{\eta^2}\right| \right.\nonumber\\ 
&\hspace{1.5 cm}\left.+ \frac{3(\eta-\eta^2-\eta^3) -1}{12\eta^3}\ln\left|\frac{1+\eta}{1-\eta}  \right|\right),
\end{align}
where $\alpha = e^2/(\kappa v_F)$ and $\eta = k_F/k_c$ with $k_c$ being the ultraviolet momentum scale beyond which the linear dispersion no longer holds. We emphasize that $k_c$ is the ultraviolet momentum scale necessary for cutting off the high-momentum logarithmic divergence associated with the linear Dirac spectrum.  For 3D Dirac systems, $k_c$ should typically be of the order of the inverse of the lattice length scale (or perhaps even a much smaller momentum scale in the realistic material where the band energy dispersion starts deviating from linearity). Using the renormalization group (RG) theory it is possible to express the interacting compressibility at one density (i.e., one $k_F$ value) simply in terms of that at another density, thus completely eliminating the unknown ultraviolet scale associated with $k_c$ as has been extensively discussed recently in Ref.~\cite{Robert15}. Note that the interaction coupling constant $\alpha = e^2/(\kappa v_F)$  is the ratio of the interaction to kinetic energy (the effective fine-structure constant), which is density independent for Dirac materials. For the case of parabolic dispersion, the coupling is density dependent and typically denoted by $r_s$ where $r_s \propto n^{-1/3}$ for regular 3D parabolic-band systems. The extrinsic self-energy for low temperatures ($t \ll 1$) can be Sommerfeld expanded, yielding 
\begin{align}
\Sigma_+^{\mathrm{ext}}(k_F) =-\frac{\alpha v_F k_F}{6\pi} \left(2 +4\ln 2 
-A\pi^2t^2  +\pi^2 t^2\ln t  \right),
\end{align}
where $A = 2\gamma + \ln (32/\pi) - 12\ln G \simeq 0.49$ with $\gamma \simeq 0.577$ being the Euler's constant and $G \simeq 1.282$ being the Glaisher's constant.

Within the HFA, the interacting chemical potential is the sum of noninteracting part and exchange self-energy at Fermi momentum:
\begin{equation}
\mu = \mu_0 + \Sigma^{\mathrm{int}}_+(k_F) + \Sigma^{\mathrm{ext}}_+(k_F).
\end{equation}
Differentiating the chemical potential with respect to the density, we then obtain the low-temperature interacting incompressibility $K_0/K(T)$ as
\begin{align}\label{eq:HFAcompress}
\frac{K_0}{K(T)} &= 1+\frac{\pi^2 t^2}{3} - \frac{\alpha}{\pi}\left\{\tilde{g}\left(\frac{k_F}{k_c}\right)+  \frac{1}{3}+\frac{2}{3}\ln 2\right.\nonumber\\
&\hspace{2 cm}\left. - \frac{\pi^2t^2}{6}(1- A) -\frac{\pi^2t^2}{6}  \ln t\right\},
\end{align}
where
\begin{align}\label{eq:gell}
\tilde{g}(\eta) &= \frac{1}{12 \eta^3} \left[6\eta^2-4\eta-4\eta^3\ln\left|1-\frac{1}{\eta^2}\right| \right.\nonumber\\
&\hspace{2 cm}\left.+ (2-3\eta-3\eta^3)\ln \left|\frac{1+\eta}{1-\eta} \right| \right] \nonumber\\
&=\frac{1}{9} + \frac{2}{3}\ln \eta + \mathcal{O}(\eta).
\end{align}
Similar to the case of graphene and parabolic-band systems~\cite{Qiuzi}, the low-temperature HF incompressibility of a 3D Dirac material shows a leading-order $t^2 \ln t$ temperature dependence followed by a next-to-leading-order $t^2$ term.

\begin{figure}[h]
\capstart
\begin{center}
\includegraphics[width=\linewidth]{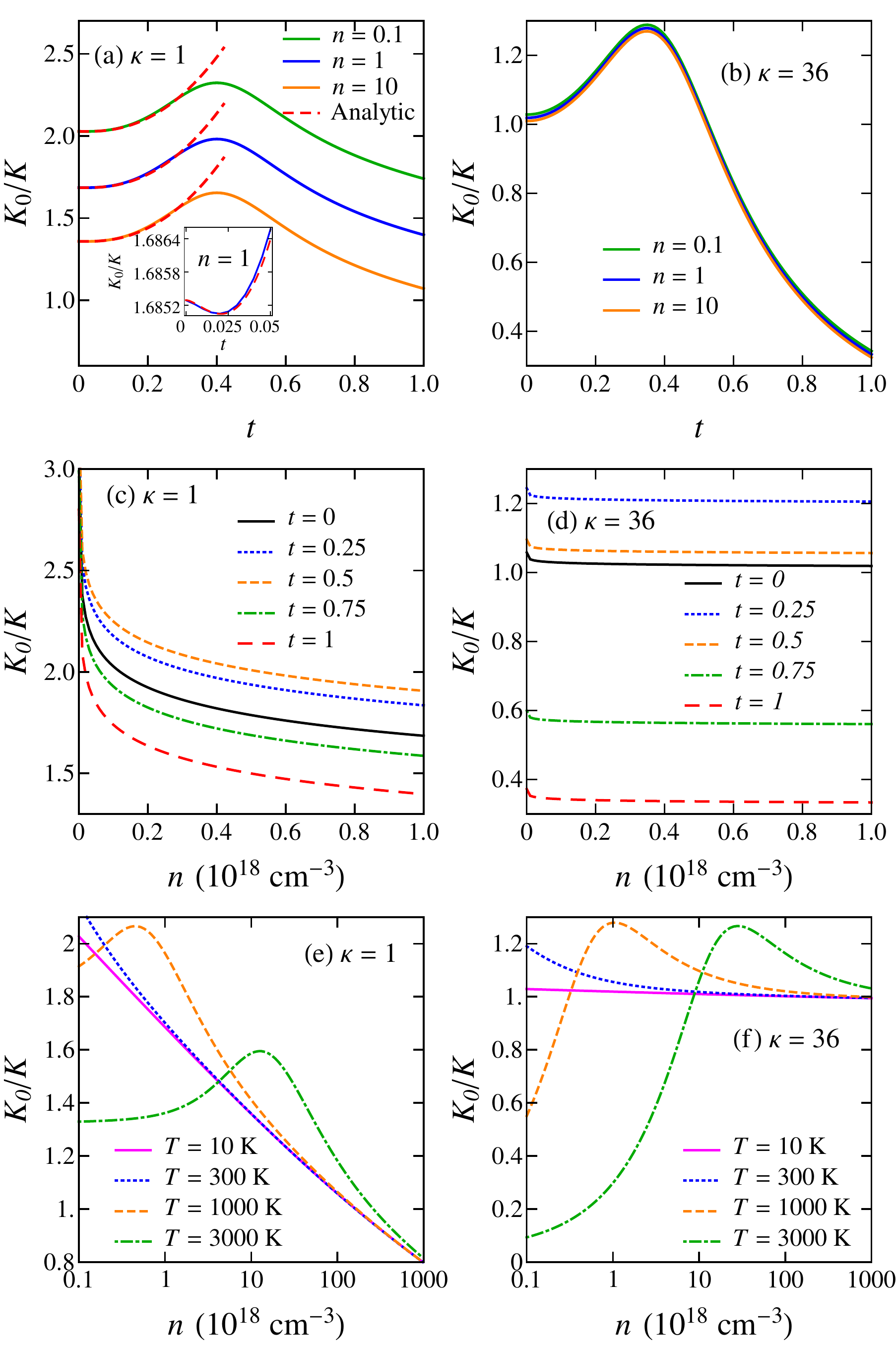}
\end{center}
\caption{(Color online) (Top) Incompressibility $K_0/K$ as a function of normalized temperature $t=T/T_F$ for three different carrier densities $n$ (expressed in the unit of $10^{18}$ cm$^{-3}$) and dielectric constants: (a) $\kappa = 1$ and (b) $36$. Inset of (a) shows the low-temperature behavior of the incompressibility. Solid lines denote the numerical results and red dashed lines denote the asymptotic low-temperature analytical results. (Middle) Incompressibility $K_0/K$ as a function of carrier density $n$ for different values of normalized temperature $t$ and dielectric constants: (c) $\kappa = 1$ and (d) $36$. (Bottom) Semilog plot of $K_0/K$ as a function of carrier density $n$ for different values of temperature $T$ and dielectric constant: (e) $\kappa = 1$ and (f) $36$. The above plots are calculated using the parameters: wave vector cutoff $k_c = 2\pi/a_c$ ($a_c$ = 16 \AA{}), degeneracy $g = 4$, and Fermi velocity $v_F = 10^8$ cm/s.}\label{fig:fig1}
\end{figure}

Figure~\ref{fig:fig1} shows the plots of the exact numerically calculated single-loop finite-temperature incompressibility $K_0/K$. The results are computed by using wave vector cutoff  $k_c = 2\pi/a_c$ ($a_c = 16 $ \AA{} is the geometric mean of the lattice constants in Cd$_3$As$_2$), degeneracy $g = 4$ and Fermi velocity $v_F = 10^{8}$ cm/s. We provide our numerical results using the parameters for Cd$_3$As$_2$ unless otherwise stated. Figs.~\ref{fig:fig1}(a) and (b) show the temperature dependence of the incompressibility $K_0/K$ for three different values of carrier density $n = 0.1$, 1 and $10$ (given in the unit of $10^{18}$ cm$^{-3}$) with dielectric constants: $\kappa = 1$ and $\kappa = 36$ (dielectric constant of Cd$_3$As$_2$ at 4.2 K)~\cite{Gerin}. As can be seen from the figures, the temperature dependence is nonmonotonic. At ``low" temperatures (i.e., $t\ll 1$) corresponding to high values of doping density (and  consequently $T_F$ can be quite large), the leading-order temperature dependence is $t^2\ln t$ where it develops a shallow minimum with the size of the minimum proportional to the value of $\alpha$. Due to the smallness of the prefactor $\alpha\pi/6$ [Eq.~\eqref{eq:HFAcompress}], this minimum is very shallow as shown in the inset of Fig.~\ref{fig:fig1}(a). The low-temperature asymptotic analytical results [Eq.~\eqref{eq:HFAcompress}], shown as red dashed lines in Fig.~\ref{fig:fig1}(a), agree precisely with the numerical results. As the temperature increases further, the incompressibility begins to increase until it reaches a maximum and beyond that it decreases with increasing temperature. Unlike the parabolic-band electron gas where the exchange-energy contribution to the compressibility vanishes at high temperatures~\cite{Qiuzi}, for Dirac materials there is a finite contribution of the exchange self-energy to the compressibility at large $t$. This interesting persistence of quantum many-body effects into the classical $T\gg T_F$ regime presumably arises from the fact, pointed out in Ref.~\cite{DasSarma09}, that Dirac systems with linear energy dispersion are manifestly nonclassical and do not have any classical analogs since a linear-in-momentum energy dispersion is manifestly quantum-mechanical and not allowed classically. We note that a large dielectric constant suppresses the electron-electron interaction strength (since the coupling constant is inversely proportional to the background dielectric constant) which weakens the density dependence of $K_0/K$ as can be seen by comparing Figs.~\ref{fig:fig1}(a) and (b).

\begin{figure}[h]
\capstart
\begin{center}
\includegraphics[width=\linewidth]{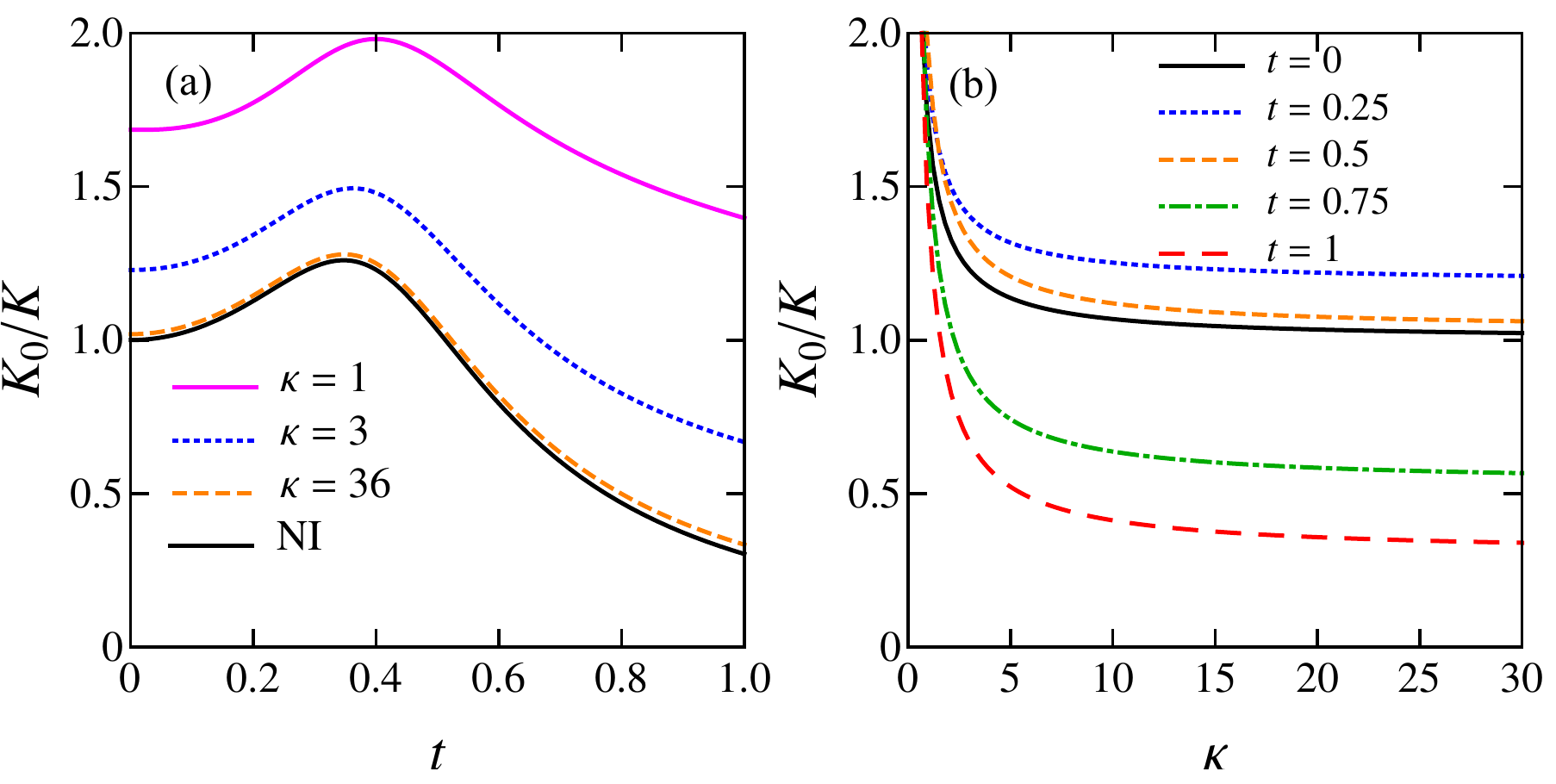}
\end{center}
\caption{(Color online) (a) Incompressibility $K_0/K$ as a function of normalized temperature $t=T/T_F$ for three different values of dielectric constant $\kappa = 1, 3$ and $36$. The black solid line (NI) denotes the noninteracting incompressibility. (b) Incompressibility $K_0/K$ as a function of dielectric constant $\kappa$ for different values of normalized temperature $t$. The plots are calculated using the parameters: $n = 10^{18}$ cm$^{-3}$, $g = 4$, $v_F = 10^8$ cm/s and $k_c = 2\pi/a_c$ ($a_c$ = 16 \AA{}).}\label{fig:fig2}
\end{figure}

\begin{figure}[h]
\capstart
\begin{center}
\includegraphics[width=\linewidth]{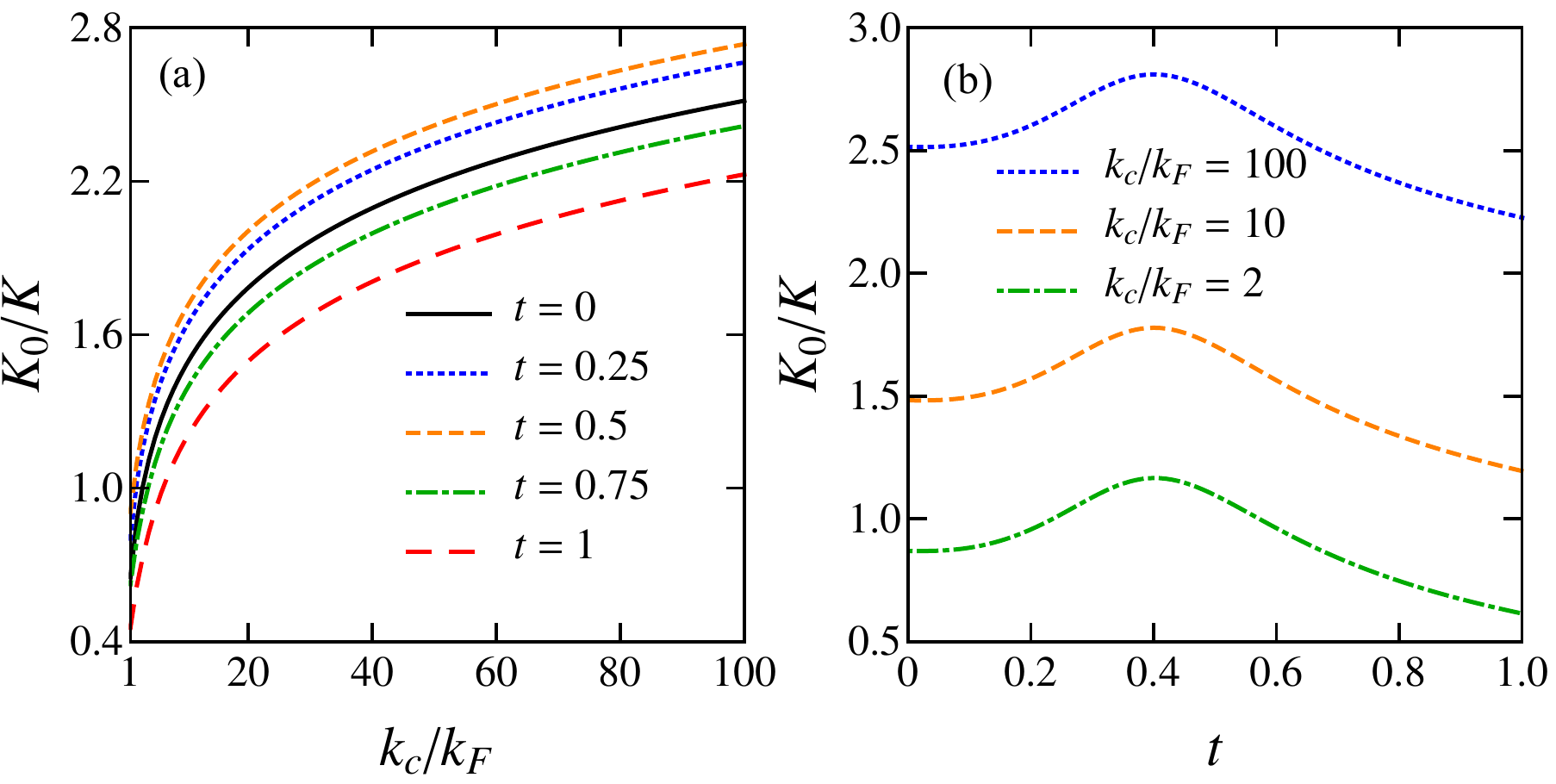}
\end{center}
\caption{(Color online) (a) Incompressibility $K_0/K$ as a function of wave vector cutoff  $k_c/k_F$ for different temperatures $t = T/T_F$. (b) Incompressibility $K_0/K$ as a function of normalized temperature $t$ for different values of wave vector cutoff $k_c/k_F$. The above plots are calculated using $\alpha = 2.2$.}\label{fig:fig3}
\end{figure}

\begin{figure}[h]
\capstart
\begin{center}
\includegraphics[width=\linewidth]{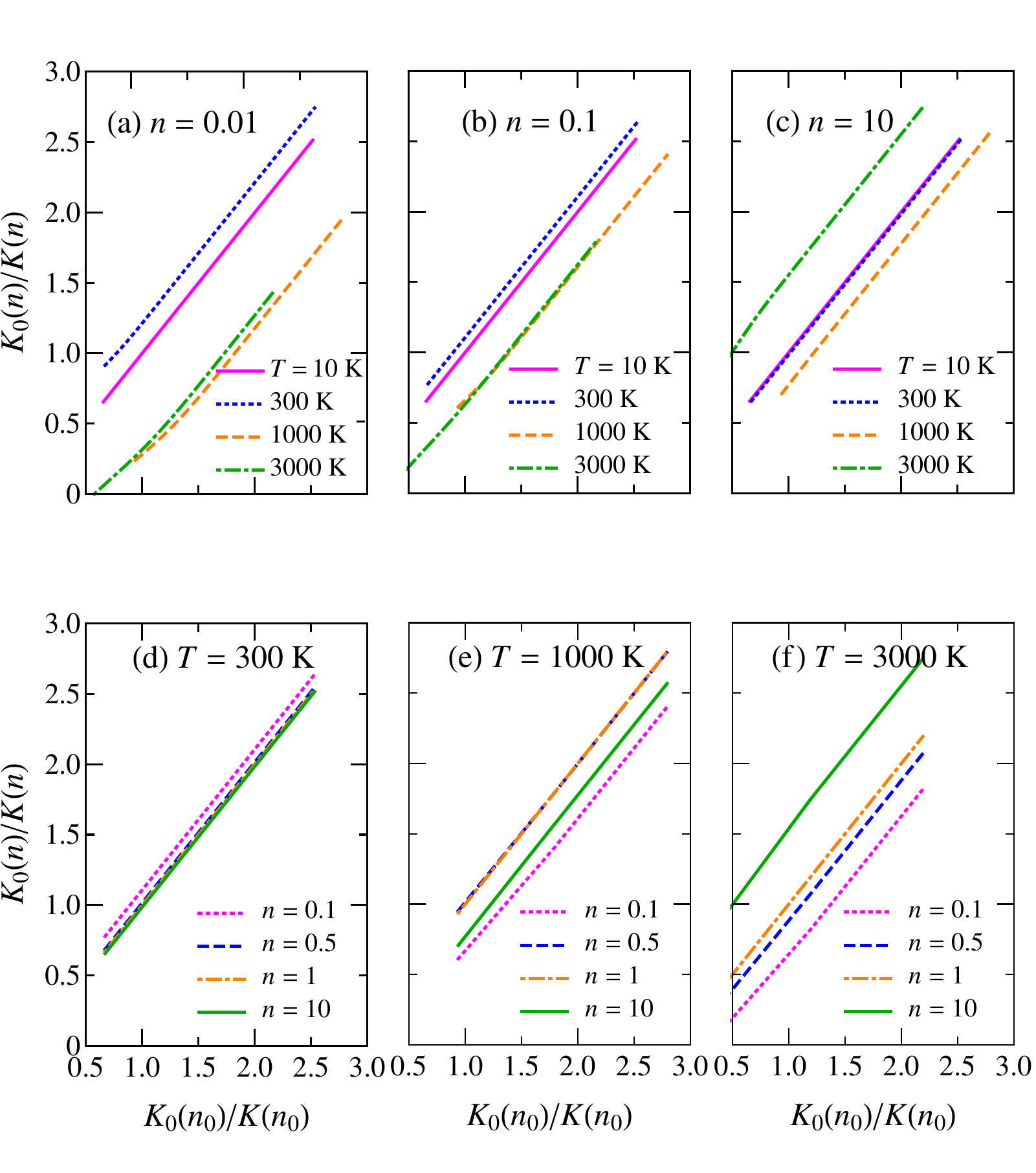}
\end{center}
\caption{(Color online) (Top) Incompressibility $K_0(n)/K(n)$ vs $K_0(n_0)/K(n_0)$ as $k_c/k_F$ is varied from 1 to 100 for different temperatures $T$ and various values of density $n$ (given in the unit of $10^{18}$ cm$^{-3}$): (a) $n = 0.01 $, (b) $0.1$, and (c) $10$. (Bottom) Incompressibility $K_0(n)/K(n)$ versus $K_0(n_0)/K(n_0)$ as $k_c/k_F$ is varied from 1 to 100 for different densities $n$ ($\times 10^{18}$ cm$^{-3}$) and various values of temperature: (d) $T = 300$ K, (e) $1000$ K, and (f) $3000$ K. In all of the above plots we set $n_0 = 10^{18}$ cm$^{-3}$ and $\alpha = 2.2$.}\label{fig:fig4}
\end{figure}

The middle panel of Fig.~\ref{fig:fig1} shows the density dependence of $K_0/K$ for several fixed values of $t$. For fixed $t$, the incompressibility decreases with the carrier density $n$ due to the decrease in the ratio of $k_c$ to $k_F$ (where $k_c$ is assumed to be fixed for a given material and $k_F$ increases with density $n$). This is to be contrasted with the case of regular 3D parabolic-band systems~\cite{Qiuzi} where for fixed $t$, the incompressibility increases with $n$. Note that a large background dielectric constant suppresses the density dependence of the incompressibility.  The nonmonotonic dependence of the incompressibility on temperature can also be seen from Figs.~\ref{fig:fig1}(c) and (d).  The bottom panel shows the semilog plot of the incompressibility $K_0/K$ as a function of carrier density $n$. For a fixed temperature $T$, at a high carrier density which corresponds to the low-$t$ limit, $K_0/K$ is a decreasing function of $n$  manifesting $n^{-2/3}$ density dependence while at a low carrier density corresponding to the high-$t$ limit, $K_0/K$ increases as $n^{2/3}$. This dependence on $n$ (where $n \propto t^{-3}$) follows mainly from the temperature dependence of the noninteracting part [Eq.~\eqref{eq:finTcomp}]. 

Figure~\ref{fig:fig2}(a) displays the temperature dependence of $K_0/K$ for different values of the background dielectric constant $\kappa$. The noninteracting part of the compressibility becomes more dominant for larger dielectric constants as the electron-electron interaction strength is suppressed in the presence of large dielectric constant. The ultraviolet momentum cutoff dependence of $K_0/K$ is shown in Fig.~\ref{fig:fig3}. As shown in the figure, the incompressibility increases with increasing momentum cutoff [with a $\ln(k_c/k_F)$ dependence for large cutoff]. To better aid experimentalists in interpreting the cutoff dependence, in Fig.~\ref{fig:fig4}, we plot $K_0(n)/K(n)$ for different values of densities $n$ against $K_0(n_0)/K(n_0)$ at $n_0 = 10^{18}$ cm$^{-3}$ as the momentum cutoff $k_c/k_F$ is varied from 1 to 100.

We note that experimentally the ultraviolet cutoff $k_c$ is of course unknown (although using a cutoff corresponding to the band momentum where the relevant band dispersion deviates from linearity or even the lattice cutoff should both be fairly decent approximations), and therefore, the interaction coupling should be considered as a running coupling in the standard RG sense as discussed in details recently in Ref.~\cite{Robert15}.  A practical way to compare theory with experiment would be to eliminate $k_c$ and express the experimental compressibility at one density with that at another density.  Taking one of these densities to be a fixed fiducial density, the measured compressibility at an arbitrary density could be expressed in terms of this fiducial compressibility and thus compared with the theory.  Such a comparison, as described already in the literature~\cite{Robert15,Barnes14}, directly tests the running coupling constant idea of quantum field theoretic RG flow in the specific context of solid-state Dirac materials.

\section{Finite Temperature renormalized quasiparticle velocity}\label{sec:velocity}
In this section we calculate the finite-temperature renormalized quasiparticle velocity for extrinsic 3D and 2D Dirac materials with $E_F > 0$ within the single-loop HFA. We note that the interaction theory for the compressibility developed in the last section implicitly incorporates many-body velocity renormalization effects, but it is useful to provide the explicit results for the quasiparticle velocity renormalization in Dirac materials as a function of temperature since the quasiparticle velocity enters as a central quantity determining many properties of Dirac systems.  There is no available theory for the temperature dependence of the quasiparticle velocity renormalization in 3D and 2D Dirac materials. Here, we provide temperature dependent results for both cases. In particular, we give the full numerical results for the one-loop HF renormalized quasiparticle velocity at $k=0$ together with the low-temperature analytic expressions for the one-loop HF renormalized quasiparticle velocity at $k = 0$ and $k = k_F$. Although the bare velocity $v_F$ in Eq.~\eqref{eq:h0} for the effective linear band energy dispersion is by definition momentum independent for Dirac systems, interaction-induced many-body effects may introduce nonlinearity or momentum dependence since the self-energy is in general momentum dependent. In Appendix~\ref{sec:zeroT} we present the detailed calculation of zero-temperature HF quasiparticle velocity for 3D Dirac materials. Within the HFA, the renormalized quasiparticle velocity for the doped (extrinsic) system with $E_F >0$ is defined by
\begin{equation}
v_{\mathrm{ext}}(k) = \left.\frac{\partial [\varepsilon_+(k') +\Sigma_+^{\mathrm{int}}(k') +\Sigma_+^{\mathrm{ext}}(k')] }{\partial k'}\right|_{k'=k}.
\end{equation}

\subsection{3D Dirac material}
\begin{figure}[h]
\capstart
\begin{center}
\includegraphics[width=\linewidth]{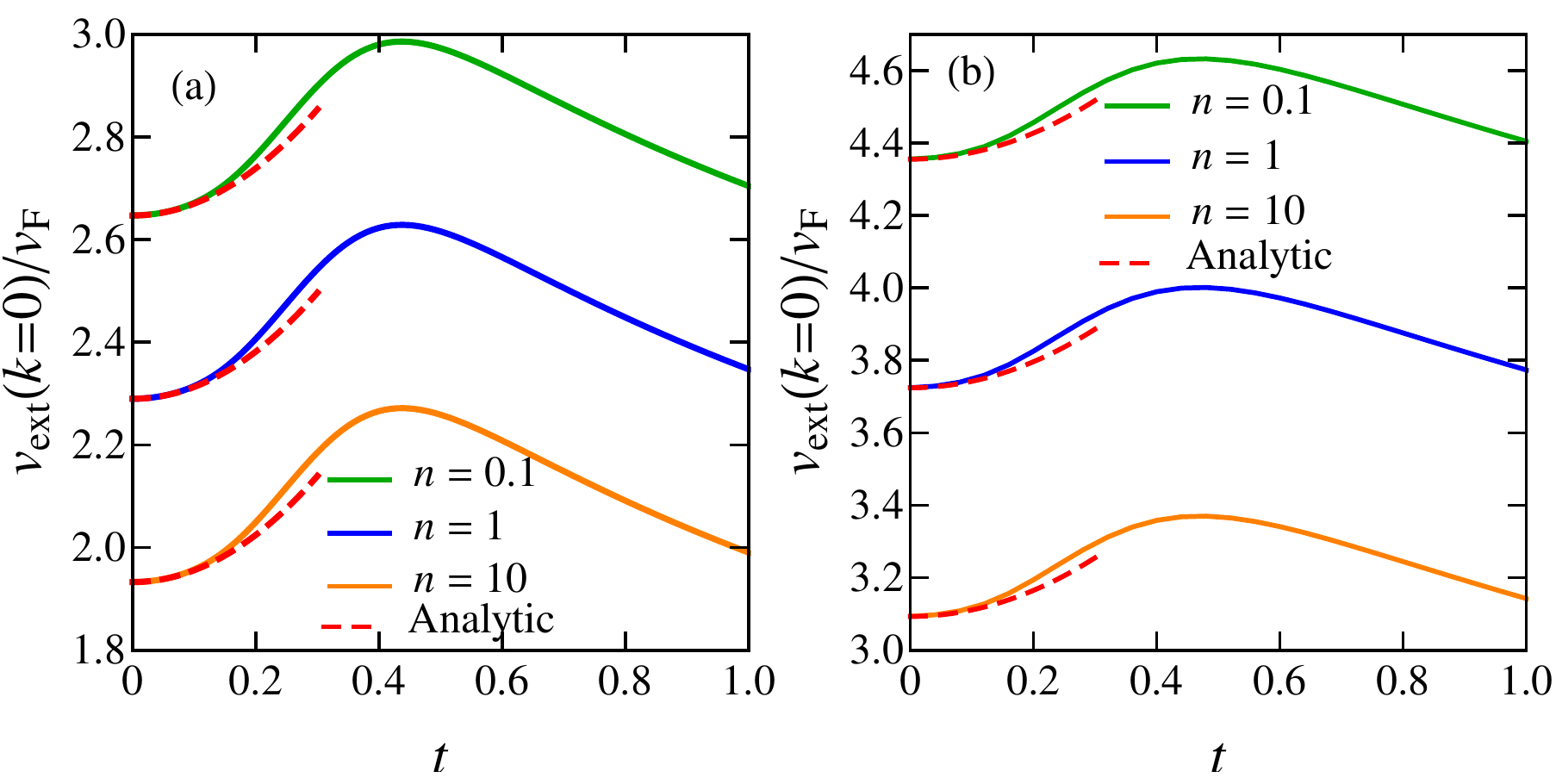}
\end{center}
\caption{(Color online) Renormalized quasiparticle velocity at $k =0$ $[v_{\mathrm{ext}}(k =0)/v_F]$ as a function of rescaled temperature $t = T/T_F$ for (a) 3D Dirac materials with wave vector cutoff $k_c = 2 \pi/a_c$ $(a_c = 16$ \AA{}) and different carrier densities $n$ $(\times 10^{18}$ cm$^{-3}$) and (b) graphene with wave vector cutoff $k_c = 2 \pi/a_c$ ($a_c = 2.46$ \AA{}) and different carrier densities $n$ $(\times 10^{12}$ cm$^{-2}$). Solid lines denote numerical results and red dashed lines denote the low-temperature analytical results. The above plots are calculated using $\alpha = 2.2$ and degeneracy $g =4$.}\label{fig:figvext}
\end{figure}

At low temperatures, the leading-order temperature correction to the renormalized HF quasiparticle velocity of a 3D Dirac semimetal at $k=0$ can be calculated to be [using Eq.~\eqref{eq:sigmaintext}]
\begin{align}\label{eq:vext}
v_{\mathrm{ext}}(k =0)=v_F \left\{ 1 + \frac{\alpha}{3\pi} \left[2\ln \left(\frac{k_c}{k_F} \right) + \pi^2t^2 \right] \right\}.
\end{align}
Thus the low-temperature renormalized quasiparticle velocity at $k = 0$ has a leading-order $t^2$ temperature dependence without any logarithmic term in temperature. Figure~\ref{fig:figvext}(a) shows the full numerical results of $v_{\mathrm{ext}}(k=0)$ for 3D Dirac materials. As shown in the figure, $v_{\mathrm{ext}}(k=0)$ has a nonmonotonic dependence with temperature where it first increases with temperature, reaches a maximum and then decreases with increasing temperature $t$. The low-temperature analytical results [Eq.~\eqref{eq:vext}], shown as red dashed lines in the figure, are in a perfect agreement with the numerical results. 

To the leading-order in $t$, the renormalized extrinsic Fermi velocity [$v^*_F = v_{\mathrm{ext}}(k_F)$] at low temperatures is
\begin{align}\label{eq:vfext}
v_F^* = v_F\left\{1 +\frac{\alpha}{\pi}\left[ 2 \ln \frac{1}{t} -\tilde{g}\left(\frac{k_F}{k_c}\right)  - B  \right] \right\},
\end{align}
where $\tilde{g}(\eta)$ is given by Eq.~\eqref{eq:gell} and $B = \frac{5}{6} - \ln({3\sqrt[3]{2}/\pi^2}) \simeq 1.79$. The $\ln (1/t)$ dependence of the renormalized extrinsic Fermi velocity implies that $v_F^* \rightarrow \infty$ as $t \rightarrow 0$. So within the HFA, the renormalized extrinsic Fermi velocity diverges logarithmically with temperature.  We expect this logarithmic divergence to disappear if the correlation effects are taken into account as, for example, in the RPA theory where the perturbative expansion is carried out in the screened Coulomb interaction~\cite{Johannes14,Johannes15a,DasSarma07}.
This logarithmic zero-temperature divergence of the renormalized Fermi velocity (obtained at $k_F$) is a well-known pathology of the HFA using the unscreened Coulomb interaction for the many-body perturbative expansion. This logarithmic HFA-induced Fermi surface divergence has been known for 3D metals for more than 50 years~\cite{Mahan,Abrikosov75}, and it happens for Dirac materials only when there is a Fermi surface, i.e., when the system is doped.  The corresponding $T=0$ logarithmic divergence in extrinsic graphene has been discussed in depth in Ref.~\cite{Hwang07}, and the same description applies here for the 3D Dirac system. We emphasize that unlike the ultraviolet logarithmic divergence associated with the momentum cutoff $k_c$ (where the divergence arises from $k_c \rightarrow \infty$) which is a defining property of the linearly dispersing Dirac-Weyl systems, the infrared logarithmic divergence arises specifically from the long-range nature of the electron-electron interaction and is generic to all electron liquids with Coulomb interaction.  This infrared divergence, being a property specific to the Fermi surface, does not happen at the band bottom ($k=0$), and therefore, there is no such infrared divergence issue for the undoped (intrinsic) Dirac system with no Fermi surface.  Thus, the velocity renormalization at $k=0$ [Eqs.~\eqref{eq:vext}] does not have the infrared divergence while the renormalized extrinsic Fermi velocity at $k_F$  [Eqs.~\eqref{eq:vfext}]  possesses the infrared divergence.  We note that, in contrast to the compressibility itself, the single-loop velocity renormalization at $k=0$ has a $T^2$ temperature dependence with no logarithmic $T^2\ln T$ leading-order correction.

\subsection{2D Dirac material}
Although the focus of the current paper is on many-body interaction effects in 3D Dirac systems, we provide in this subsection our HF single-loop results for the temperature-dependent many-body renormalization of 2D Dirac quasiparticle velocity, which is the finite-temperature generalization of the existing $T=0$ velocity renormalization results for graphene~\cite{Hwang07,DasSarma07}. We refer to the 2D Dirac system as graphene with no loss of generality because of the vast existing literature on graphene~\cite{Neto09,DasSarma11,Adam07,Kotov12}.

Solving Eq.~\eqref{eq:conserv}, we obtain the noninteracting chemical potential for graphene in the low- and high-temperature limits as~\cite{Qiuzi}
\begin{align}
\frac{\mu_0}{E_F} =
\begin{dcases}
1 - \dfrac{\pi^2 t^2}{6}, & \mbox{for } t \ll 1,\\
\dfrac{1}{4 t \ln 2}, & \mbox{for } t \gg 1.
\end{dcases}
\end{align}
Using the above chemical potential, we can then calculate the leading-order temperature correction to the low-temperature renormalized quasiparticle velocity for graphene at $k =0$ to be
\begin{equation}
v_{\mathrm{ext}}(k =0) = v_F \left\{1 + \frac{\alpha}{4} \left[ \ln\left(\frac{k_c}{k_F}\right) + \frac{\pi^2 t^2}{3} \right]\right\}.
\end{equation}  
Similar to the 3D Dirac case [Eq.~\eqref{eq:vext} above], $v_{\mathrm{ext}}(k=0)$ for graphene also increases as $t^2$. The full numerical results for $v_{\mathrm{ext}}(k=0)$ in graphene are presented in Fig.~\ref{fig:figvext}(b). The renormalized quasiparticle velocity $v_{\mathrm{ext}}(k=0)$ for graphene has a similar nonmonotonic temperature dependence to that for 3D Dirac materials where it increases as $t^2$ at low temperatures and decreases with temperature at high temperatures, reaching a maximum at some intermediate value of temperature.

To the leading order in temperature $t$, the low-temperature renormalized extrinsic Fermi velocity is
\begin{equation}
v_F^* = v_F \left\{1+ \alpha\left[\frac{2}{\pi}\ln \frac{1}{t} -p\left(\frac{k_F}{k_c} \right) - D \right]  \right\},
\end{equation}
where
\begin{subequations}
\begin{align}
p(\eta) &= \frac{1}{\pi} \left\{ \frac{(\eta -1)}{\eta^2}[E(\eta) - K_1(\eta)] -\frac{\pi}{4}\left[\ln\left(\frac{4}{\eta}\right)-\frac{1}{2} \right] \right.\nonumber\\
&\hspace{1.5 cm}+\left.\int_0^{\eta}dy\frac{1}{y^3}\left[K_1(y) - E(y) - \frac{\pi}{4} y^2\right]  \right\}\nonumber\\
&= \frac{1}{4}\left[\frac{3}{2} - \ln \left(\frac{4}{\eta}\right) \right]  + \mathcal{O}(\eta),\\
D &= \frac{1}{2\pi}\left( 3 +2C - 8\ln 2 + 2 \ln \frac{\pi^2}{3}\right) \approx 0.266,
\end{align}
\end{subequations}
with $C\approx 0.916$ being the Catalan's constant, $K_1(x)$ and $E(x)$ being the complete elliptic integrals of the first and second kind. Similar to the 3D case, there is a logarithmic dependence of the extrinsic Fermi velocity on temperature within the HFA which is exactly the same infrared divergence associated with the long-range Coulomb interaction discussed before. This infrared logarithmic divergence at $k_F$ for the extrinsic system will vanish if the correlation effects are taken into account (e.g., RPA) as for the case of zero temperature~\cite{DasSarma07,Johannes14}.

\section{Summary}\label{sec:summary}
In conclusion, we have theoretically obtained the finite-temperature interacting compressibility for extrinsic 3D Dirac materials and renormalized quasiparticle velocities for extrinsic 2D and 3D Dirac materials within the single-loop HFA using the long-range Coulomb interaction between the electrons. We give the numerical results of the incompressibility $K/K_0$ and also its analytic expression at low temperatures. The HF incompressibility $K/K_0$ has a nonmonotonic dependence with temperature $t = T/T_F$. At low temperatures, as the temperature increases the incompressibility first decreases slowly with a $t^2 \ln t$ dependence to a minimum. After reaching the minimum, as the temperature is raised further it rises to a maximum and subsequently decreases with temperature at high temperatures. We note that our HFA results are the same as the single-loop (in the bare interaction) results, and our results are therefore \textit{exact} to the leading order in the effective coupling constant.
We also present the numerical results and low-temperature analytic expression for the finite-temperature renormalized quasiparticle velocities at $k=0$ [$v_{\mathrm{ext}}(k =0)$] of 3D and 2D Dirac materials. The renormalized quasiparticle velocity at $k=0$ also has a nonmonotonic dependence on temperature, i.e., it first increases with temperature and beyond a certain value of temperature, it decreases with increasing temperature. At low temperatures the leading-order correction to $v_{\mathrm{ext}}(k =0)$ is $t^2$. Our theoretical predictions can be directly verified in the experiments through either temperature-dependent measurements at a finite doping density (with a fixed $T_F=E_F/k_B$) or through density-dependent measurements at a finite temperature since the relevant dimensionless variable determining the temperature dependence is $t=T/T_F$, where $T_F \propto n^{1/3}$ and $T_F \propto n^{1/2}$ in 3D and 2D Dirac materials, respectively.

Besides the finite-temperature results, we also discuss the zero-temperature density-dependent HF and RPA compressibilities in the Appendices. As shown in Appendix~\ref{sec:zeroT}, the HF incompressibilities for 2D and 3D Dirac materials increase logarithmically with the momentum cutoff. On the other hand, the RPA incompressibility (given in Appendix~\ref{sec:zeroRPA}) for 3D Dirac materials has a log-log dependence on the cutoff while the RPA incompressibility for 2D Dirac materials increases logarithmically with the momentum cutoff but with a smaller prefactor compared to its HF counterpart. We note that for weak coupling, where the single-loop theory is valid, both HFA and RPA give quantitatively identical results (except for the infrared logarithmic divergence for the renormalized Fermi velocity at $k=k_F$ which occurs in HFA, but not in RPA), and therefore, a systematic experiment to measure the compressibility in one of the newly discovered 3D Dirac materials, most of which have rather low values of interaction coupling constant (typically, $\alpha\sim 0.1$), as a function of density and temperature should lead to direct experimental observation of many-body effects in Dirac systems as described in this work.

\begin{acknowledgments}
This work is supported by LPS-MPO-CMTC and JQI-NSF-PFC. F.~S. thanks E. Barnes, J. Hofmann and Q. Li for enlightening discussions. F.~S. is also grateful to J. Hofmann for his instructive notes on the zero-temperature self-energy.
\end{acknowledgments}

\appendix
\section{Zero-temperature HF compressibility}\label{sec:zeroT}
In this Appendix, we give detailed calculations of the zero-temperature interacting compressibility within the HFA. The zero-temperature intrinsic and extrinsic exchange self-energies are given by~\cite{Johannes15a}
\begin{subequations}
\begin{align}
\Sigma_s^{\mathrm{int}}(k) &= - \frac{\alpha v_Fk_c}{\pi} \left[ f\left(\frac{k}{k_c} \right) - sh\left( \frac{k}{k_c} \right)  \right], \\
\Sigma_s^{\mathrm{ext}}(k) &= -\frac{\alpha v_Fk_F}{\pi} \left[f\left(\frac{k}{k_F}\right)+sh\left(\frac{k}{k_F}\right)\right],
\end{align}
\end{subequations}
where
\begin{subequations}
\begin{align}\label{eq:fxhx}
f(x) &= \frac{1}{2} + \frac{1-x^2}{4x}\ln \left|\frac{1+x}{1-x} \right| = 1 - \frac{x^2}{3} + \mathcal{O}(x^4),\\
h(x) &= -\frac{1}{6x} + \frac{x}{3}\ln\left|1-\frac{1}{x^2}\right|+ \frac{3x^2+1}{12x^2}\ln\left|\frac{1+x}{1-x}\right|\nonumber\\
& = \frac{2}{3}x \left(\frac{5}{6} - \ln x\right) + \mathcal{O}(x^3). 
\end{align}
\end{subequations}
Note that at $T=0$, the HF self-energy for a parabolic-band 3DEG is given by $-\frac{2e^2k_F}{\pi\kappa} f(\frac{k}{k_F})$.

\subsubsection{Intrinsic (undoped) Dirac material}
Near the Weyl node ($k\ll k_c$), the intrinsic self-energy can be expanded as
\begin{equation}
\Sigma_s^{\mathrm{int}}(k) = - \frac{\alpha v_Fk_c}{\pi } \left\{1-s\frac{2k}{3k_c}\left[\ln \frac{k_c}{k}+\frac{5}{6} + \mathcal{O}\left(\frac{k}{k_c}\right) \right] \right\} .
\end{equation}
The first term $-\alpha v_F k_c/\pi$ introduces a trivial shift in the energy while the remaining terms renormalize the quasiparticle velocity. The renormalized quasiparticle velocity for small $k$ $(k\ll k_c)$ is given by
\begin{align}
v_{\mathrm{int}}(k) &= \frac{\partial[ \varepsilon_s(k) + \Sigma_s^{\mathrm{int}}(k)]}{\partial k} \nonumber\\
&= v_F \left [1 + \frac{2\alpha}{3\pi}\ln \left(\frac{\widetilde{k}_c}{k} \right)+ \mathcal{O}\left(\frac{k}{k_c}\right) \right],
\end{align}
where $\widetilde{k}_c = e^{-1/6}k_c\approx 0.846 k_c$. This renormalization is also obtained in Refs.~\cite{Johannes15a,Rosenstein13,Gonzalez14,Robert15,Isobe12,Isobe13}. The logarithmic divergence of $v_{\mathrm{int}}(k)$ as $k\rightarrow 0$ is found only in the intrinsic case as for the extrinsic case this divergence is canceled out by a similar divergence in $\Sigma^{\mathrm{ext}}(k)$ as discussed below. 

\subsubsection{Extrinsic (doped) Dirac material}
For $k\ll k_F$, the extrinsic self-energy can be written as
\begin{align}
\Sigma_s^{\mathrm{ext}}(k) &= -\frac{\alpha v_Fk_F}{\pi} \left[1 + s\frac{2k}{3k_F}\left(\ln \frac{k_F}{k}+\frac{5}{6} \right) + \mathcal{O}\left(\frac{k}{k_F}\right)\right].
\end{align}
Note that since the $\ln k$ term in $\Sigma^{\mathrm{ext}}_s(k\rightarrow 0)$ cancels the same term in $\Sigma^{\mathrm{int}}_s(k\rightarrow 0)$, the renormalized quasiparticle velocity in the extrinsic case $v_{\mathrm{ext}}(k=0) = v_F [1 +\frac{2 \alpha}{3\pi}\ln (k_c/k_F)]$ has no logarithmic divergence for $k\rightarrow 0$. This cancellation of logarithmic divergence is also found for the case of graphene~\cite{Hwang07}, i.e., 2D Dirac systems. 

\begin{figure}[h]
\capstart
\begin{center}
\includegraphics[scale=0.75]{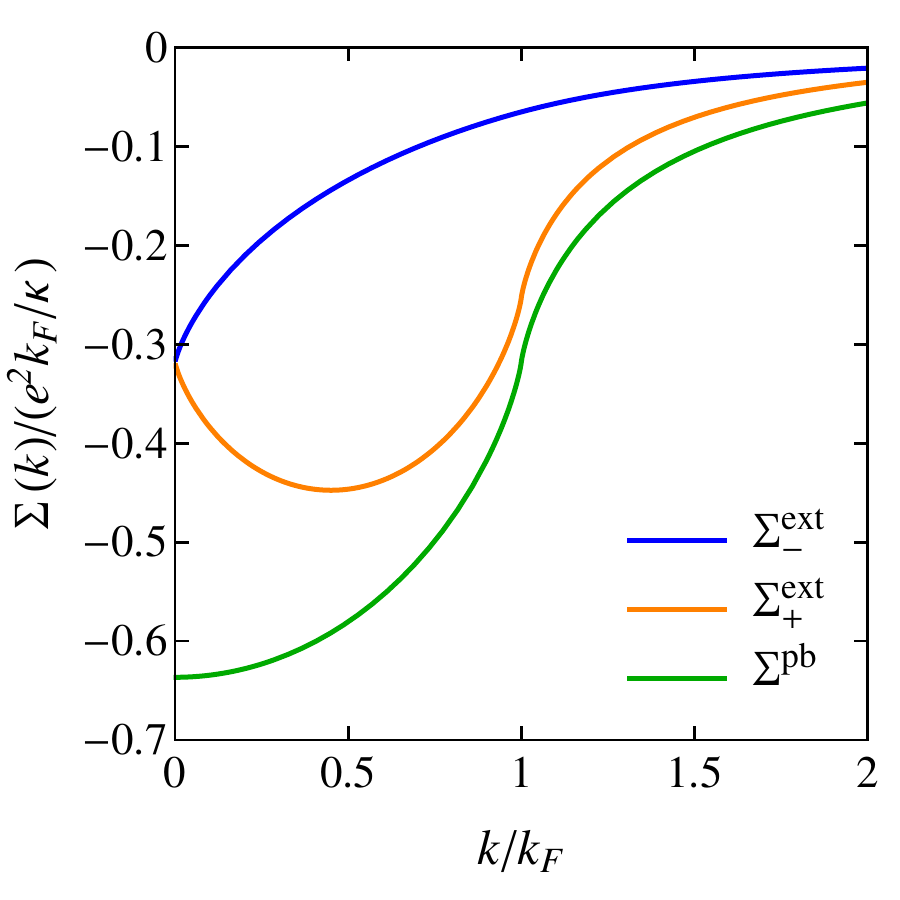}
\end{center}
\caption{(Color online) Exchange self-energies for a 3D Dirac semimetal (with $E_F >0$) and for a parabolic-band 3DEG as functions of wave number $k/k_F$. Note that $\Sigma^{\mathrm{pb}}(k) = \Sigma^{\mathrm{ext}}_+(k) + \Sigma^{\mathrm{ext}}_-(k)$. }\label{fig:F1}
\end{figure}

At the Fermi wave vector $k_F$, the extrinsic self-energy is
\begin{align}
\Sigma_s^{\mathrm{ext}}(k_F) &= -\frac{\alpha v_Fk_F}{\pi} \left[\frac{1}{2} + s\left( \frac{2}{3}\ln 2 - \frac{1}{6}\right)\right].
\end{align}
So compared to the self-energy of a parabolic-band 3DEG, the extrinsic self-energy can be written as
\begin{equation}
\Sigma^{\mathrm{ext}}_s(k_F) = \left[\frac{1}{2} + s\left( \frac{2}{3}\ln 2 - \frac{1}{6}\right)\right]  \Sigma^{\mathrm{pb}}(k_F),
\end{equation}
where $\Sigma^{\mathrm{pb}}(k_F) = - e^2k_F/(\kappa\pi)$ is the self-energy of a parabolic-band 3DEG at Fermi momentum. Figure~\ref{fig:F1} shows the exchange self-energies as functions of wave number $k/k_F$ for a 3D Dirac material with $E_F > 0$. As for $\Sigma^{\mathrm{pb}}$, there is a logarithmic divergence in the slope of $\Sigma^{\mathrm{ext}}_+$ as $k \rightarrow k_F$ which manifests as the divergence in the renormalized extrinsic Fermi velocity. This is the well-known infrared HFA divergence associated with the unscreened Coulomb interaction generic to all electron liquids, which disappears in higher-order approximations, e.g., RPA. However, this singularity does not have any pathological effect on $\partial \mu/\partial n$. On the other hand, $\Sigma^{\mathrm{ext}}_-$ has a finite slope at $k =k_F$ since for extrinsic 3D Dirac material with $E_F > 0$ there is no Fermi surface at $k = k_F$ in the valence band.

\begin{figure}[h]
\capstart
\begin{center}
\includegraphics[scale=0.75]{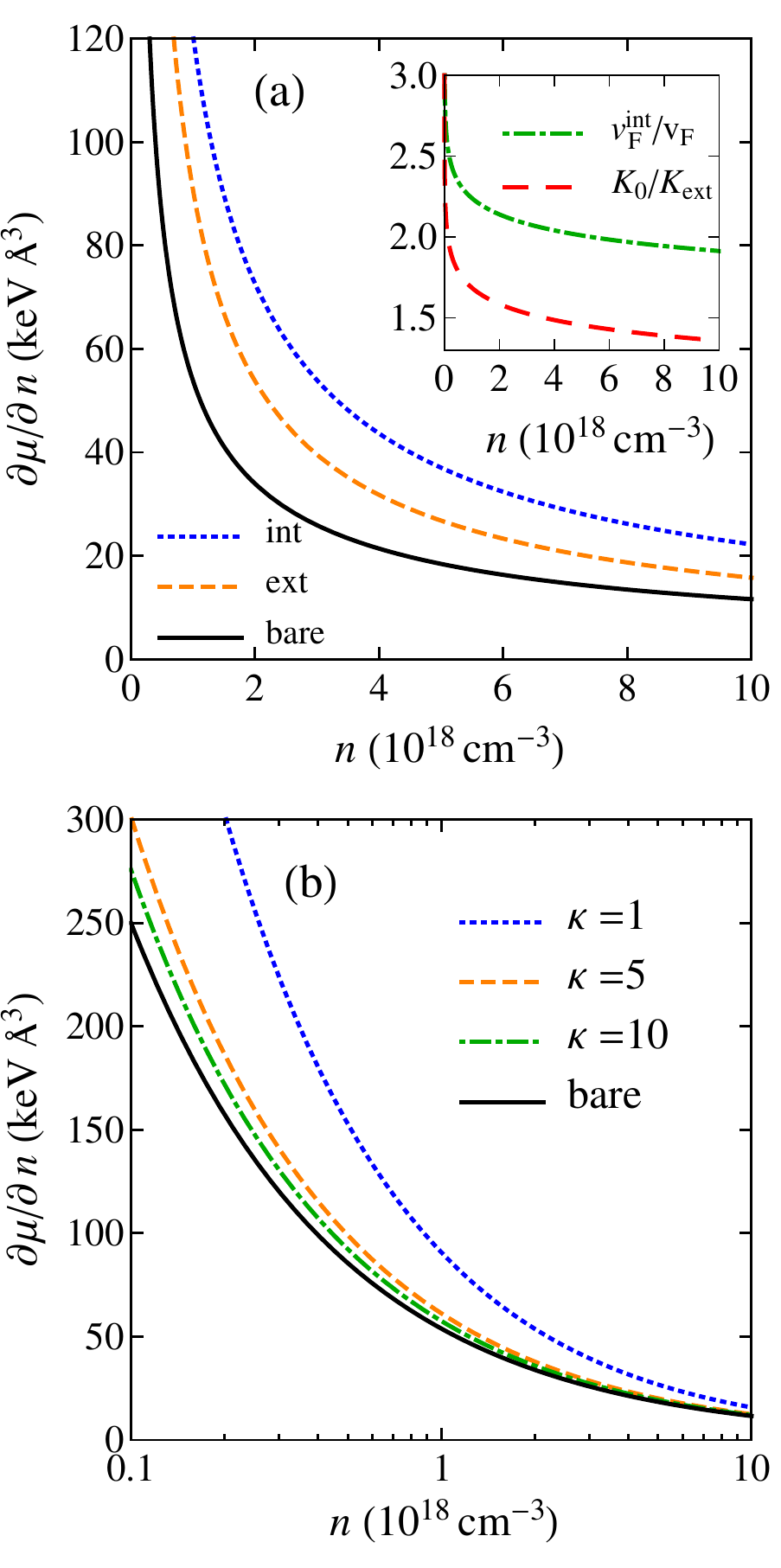}
\end{center}
\caption{(Color online) (a) Plot of $\partial \mu/\partial n$ as a function of carrier density $n$ calculated using the parameters: $\kappa = 1$, $v_F = 10^8$ cm/s, $k_c = 2\pi/a_c$ ($a_c = 16 $ \AA{}), $g = 4$, and $\alpha = 2.2$. The notations ``bare",``int", and ``ext" denote the noninteracting, intrinsic and extrinsic quantities. Inset: Plot of renormalized intrinsic velocity at $k_F$ and extrinsic incompressibility as a function of density $n$. (b) Semilog plot of $(\partial \mu/\partial n)_{\mathrm{ext}}$ as a function of carrier density $n$ for different dielectric constants $\kappa$. The noninteracting $\partial \mu_0/\partial n$ is shown as a black solid line for a comparison.}\label{fig:F2}
\end{figure}

For doped 3D Dirac materials with $E_F > 0$, the interacting chemical potential within the HFA is given by
\begin{equation}
\mu = \mu_0 + \Sigma^{\mathrm{int}}_+(k_F) + \Sigma^{\mathrm{ext}}_+(k_F).
\end{equation}
Taking the derivative of the chemical potential with respect to the free carrier density and ignoring terms of order $k_F/k_c$, we have
\begin{align}
\left(\frac{\partial\mu}{\partial n}\right)_{\mathrm{ext}} &= \frac{\partial \mu_0}{\partial n} \left[1-\frac{\alpha}{\pi}\left(\frac{4}{9} + \frac{2}{3}\ln \frac{2k_F}{k_c} \right) \right] \nonumber\\
&= \left(\frac{\partial\mu}{\partial n}\right)_{\mathrm{int}}\left[ 1 -\frac{\alpha_{\mathrm{int}}}{\pi} \left(\frac{1}{3} + \frac{2}{3} \ln 2\right)\right],
\end{align}
where $\partial \mu_0/\partial n = v_F[2\pi^2/(9n^2 g)]^{1/3}$, $\left(\frac{\partial \mu}{\partial n}\right)_{\mathrm{int}} = v^{\mathrm{int}}_F[2\pi^2/(9n^2 g)]^{1/3}$, and $\alpha_{\mathrm{int}} = e^2/(\kappa v_F^{\mathrm{int}})$ [$v_F^{\mathrm{int}} \equiv v_{\mathrm{int}}(k_F)$]. The exchange interaction changes $\partial \mu/\partial n$ from its noninteracting value $\partial \mu_0/\partial n$ by $-\frac{\alpha}{\pi}[\frac{4}{9} + \frac{2}{3}\ln (\frac{2k_F}{k_c})]$ and from its intrinsic value $(\partial \mu/\partial n)_{\mathrm{int}}$ by $-\frac{\alpha_{\mathrm{int}}}{\pi}(\frac{1}{3} + \frac{2}{3}\ln 2)$. So for carrier density $n = 5 \times 10^{18}$ cm$^{-3}$, degeneracy $g=4$, and $k_c = 2 \pi/a_c$ ($a_c = 16$ \AA{}), $(\partial \mu/\partial n)_{\mathrm{ext}}$ is increased from its bare value by $\approx 0.333 \alpha$ but decreased from its intrinsic value by $\approx -0.253 \alpha_{\mathrm{int}}$. Figure~\ref{fig:F2}(a) shows explicitly the comparison between the noninteracting, intrinsic and extrinsic $\partial \mu/\partial n$. Using dielectric constant $\kappa = 36$ (for Cd$_3$As$_2$)~\cite{Gerin} and $v_F = 10^8$ cm/s, the value of $\alpha$ can be calculated to be $\approx 0.06$ and $\partial \mu/\partial n$ is increased by roughly 2\% over its bare value. The effect of dielectric constant $\kappa$ on $(\partial \mu/\partial n)_{\mathrm{ext}}$ is shown in Fig.~\ref{fig:F2}(b). Larger dielectric constant suppresses the electron-electron interaction strength which in turn decreases the value of $\partial \mu/\partial n$. For the parameters: $\kappa = 1$, $v_F = 10^8$ cm/s and $n = 10^{18}$ cm$^{-3}$, $(\partial \mu/\partial n)_{\mathrm{ext}}$ is increased from its bare value by about 70\%.

At zero temperature, the incompressibility of an extrinsic 3D Dirac material can be calculated to be
\begin{align}
\frac{K_0}{K} &= 1-\frac{\alpha}{\pi}\left[\tilde{g}(\eta)+\frac{1}{3}+\frac{2}{3}\ln 2\right],
\end{align}
where $\tilde{g}(\eta)$ is given in Eq.~\eqref{eq:gell} and $\eta = k_F/k_c$.
On the other hand, within the HFA, the zero-temperature incompressibility for a parabolic-band 3DEG is
\begin{equation}
\frac{K_0}{K^{\mathrm{pb}}} = 1-\frac{r_s}{\pi}\left(\frac{4}{9\pi}\right)^{1/3},
\end{equation}
where $r_s = (3/4\pi n)^{1/3}(me^2\kappa) \propto n^{-1/3}$. So $K_0/K^{\mathrm{pb}}$ increases with the carrier density $n$. This is contrary to the compressibility of a 3D Dirac material, where the compressibility decreases with $n$.

\section{Zero-temperature RPA compressibility}\label{sec:zeroRPA}
In this Appendix, we calculate the leading-order ultraviolet divergent piece of the zero-temperature RPA compressibility for 2D and 3D extrinsic Dirac materials. The RPA self-energy can be written as
\begin{align}\label{eq:RPA}
\Sigma_s(i\omega_n,\mathbf{k}) = -\frac{1}{\beta V}& \sum_{\mathbf{q}s'\Omega_m} G_{s'}(i\omega_n+i\Omega_m,\mathbf{k}') \nonumber\\
&\times\left[\frac{V_{\mathbf{q}}}{\epsilon(i\Omega_m,\mathbf{q})}\right] F_{ss'}(\mathbf{k},\mathbf{k}'),
\end{align}
where $\beta = 1/(k_B T)$, $\mathbf{k}' = \mathbf{k}+\mathbf{q}$ and $\epsilon(i\Omega_m,\mathbf{q}) = 1 + V_\mathbf{q}\Pi(i\Omega_m,\mathbf{q})$ is the frequency-dependent dielectric constant with the polarizability given by
\begin{equation}
\Pi(i\Omega_m,\mathbf{q}) = - \frac{g}{V}\sum_{\mathbf{k}ss'}\frac{n_F(\xi_{s\mathbf{k}})-n_F(\xi_{s'\mathbf{k}'})}{i\Omega_m+\xi_{s\mathbf{k}} - \xi_{s'\mathbf{k}'}}F_{ss'}(\mathbf{k},\mathbf{k}').
\end{equation}
The RPA self-energy is obtained by replacing the ``bare" interaction by dynamically screened interaction (i.e., geometric series over polarization bubbles with the ``bare" interaction)~\cite{Mahan,Giuliani,Abrikosov75}. The sum of the series removes the infrared divergence of the `bare' Coulomb interaction.
Following Ref.~\cite{Quinn}, we can write down the self-energy as a sum of line and residue term:
\begin{subequations}
\begin{align}
\Sigma_s(\omega,\mathbf{k}) &= \Sigma_s^{\mathrm{line}}(\omega,\mathbf{k})+ \Sigma_s^{\mathrm{res}}(\omega,\mathbf{k}),\\
\Sigma_s^{\mathrm{line}}(\omega,\mathbf{k})&= -\frac{1}{V}\sum_{\mathbf{q}s'}\int\frac{d\Omega}{2\pi}G_{s'}(\omega+i\Omega,\mathbf{k}')\nonumber\\
&\hspace{1 cm}\times\left[\frac{V_{\mathbf{q}}}{\epsilon(i\Omega,\mathbf{q})} \right]F_{ss'}(\mathbf{k},\mathbf{k}'),\\
\Sigma_s^{\mathrm{res}}(\omega,\mathbf{k}) &= \frac{1}{V}\sum_{\mathbf{q}s'}[\Theta (\omega-\xi_{s'\mathbf{k}'}) - \Theta (-\xi_{s'\mathbf{k}'})]\nonumber\\
&\hspace{1 cm}\times\left[\frac{V_{\mathbf{q}}}{\epsilon(\xi_{s'\mathbf{k}'}-\omega,\mathbf{q})} \right]F_{ss'}(\mathbf{k},\mathbf{k}'),
\end{align}
\end{subequations}
where $\Sigma_s^{\mathrm{line}}$ is obtained by carrying out the analytic continuation $i\omega_n \rightarrow \omega +i0$ in Eq.~\eqref{eq:RPA} before performing the Matsubara sum and the residue term gives the correction to the self-energy due to the noncommutativity of these two operations. Using dimensionless energy ($\hat{\Sigma} = \Sigma/E_F$, $\nu = \omega/E_F$, $\hat{\Omega} = \Omega/E_F$) and dimensionless momenta ($x = k/k_F$, $x' = k'/k_F$, $y = q/k_F$), we can write the self-energy as
\begin{widetext}
\begin{subequations}
\begin{align}
\hat{\Sigma}_s^{\mathrm{line}}(\nu,\mathbf{x}) &= -\alpha \sum_{s'}\int_0^{\infty} dy \int \frac{d\Omega_d}{(2\pi)^d}\Omega_d F_{s s'}(\mathbf{x},\mathbf{x}')\int \frac{d\hat{\Omega}}{2\pi} \frac{1}{\epsilon(i\hat{\Omega},\mathbf{y})} \frac{1}{\nu+i\hat{\Omega}-s'|\mathbf{x}'|+1},\\
\hat{\Sigma}_s^{\mathrm{res}}(\nu,\mathbf{x}) &= -\alpha \sum_{s'}\int_0^{\infty} dy \int \frac{d\Omega_d}{(2\pi)^d}\Omega_d F_{s s'}(\mathbf{x},\mathbf{x}')\frac{[\Theta(\nu+1-s'|\mathbf{x}'|) - \Theta(1-s'|\mathbf{x}'|)]   }{\epsilon(s'|\mathbf{x}'| -1 -\nu,\mathbf{y})},
\end{align}
\end{subequations}
where $\alpha = e^2/(\kappa v_F)$ and $\Omega_d$ is the solid angle element of dimension $d$.

The chemical potential is determined by the electron energy at $k = k_F$, i.e.,
\begin{equation}
\frac{\mu}{E_F} = \frac{\mu_{0}}{E_F} + \hat{\Sigma}_{s}(0,\mathbf{1}).
\end{equation}
For the self-energy at $\nu = 0$ and $x =1$, the residue term is zero and hence $\hat{\Sigma}_{s}(0,\mathbf{1}) = \hat{\Sigma}_s^{\mathrm{line}}(0,\mathbf{1})$. So we are left with evaluating only $\hat{\Sigma}_s^{\mathrm{line}}(0,\mathbf{1})$. The calculation of the self-energy is regulated by using momentum and frequency cutoff $\ell = k_c/k_F$: 
\begin{align}
\hat{\Sigma}_s(0,\mathbf{1}) &= -\alpha \int_0^{\ell} dy  \int_{0}^{\ell}\frac{d\hat{\Omega}}{2\pi} \frac{1}{\epsilon (i\hat{\Omega},\mathbf{y})}\sum_{s'}\int \frac{d\Omega_d}{(2\pi)^d}\Omega_d \frac{F_{ss'}(\mathbf{1},\mathbf{x}')}{i\hat{\Omega} - s'|\mathbf{x}'|+1}.
\end{align}
Since the ultraviolet divergence stems from the integration over high-momentum region, we introduce the following parametrization
\begin{subequations}
\begin{align}
y &= P \sin \chi, \\
\hat{\Omega} &= P \cos \chi,
\end{align}
\end{subequations}
where $0 \leq \chi \leq \pi$, and evaluate the divergence part of the self-energy by focusing on the large-$P$ region. In this region, we have
\begin{align}\label{eq:prop}
&\sum_{s'}\int \frac{d\Omega_d}{(2\pi)^d}\Omega_d \frac{F_{ss'}(\mathbf{1},\mathbf{x}')}{i\hat{\Omega} - s' |\mathbf{x}'| + 1} = \begin{dcases}
-\frac{i}{P}\cos {\chi} +\frac{2}{\pi P^2}\left[\cos 2\chi -\frac{s}{3}(2+\cos 2\chi)\right] + \mathcal{O}\left(\frac{1}{P^3}\right)&\mbox{(3D)},\\
-\frac{i}{P}\cos {\chi} + \frac{1}{P^2}\left[\cos 2\chi - s \cos^2 \chi \right] + \mathcal{O}\left(\frac{1}{P^3}\right)&\mbox{(2D)}.
\end{dcases}
\end{align}
\end{widetext}
Since the term of order $\mathcal{O}(1/P)$ is odd in $\cos \chi$, it will be zero after doing the integral over $\chi$. It remains then to evaluate the term $\mathcal{O} (1/P^2)$. Since we are interested only in the divergence part of the self-energy, in the self-energy calculation we use the intrinsic dielectric function which is given by 
\begin{equation}\label{eq:dielec}
\epsilon(i\hat{\Omega},\mathbf{y}) = 
\begin{dcases}
1 + \dfrac{ g \alpha}{6\pi} \ln \dfrac{\ell^2}{y^2+\hat{\Omega}^2} = 1+ \lambda \ln \dfrac{\ell}{P} & \mbox{(3D)}, \\
1+ \dfrac{g \pi \alpha}{8} \dfrac{y}{\sqrt{y^2 + \hat{\Omega}^2}} =  1+ \lambda \sin \chi & \mbox{(2D),}
\end{dcases}
\end{equation}
where 
\begin{align}
\lambda = 
\begin{dcases}
\frac{g \alpha}{3\pi}, &\mbox{(3D)}, \\
\frac{g \pi\alpha}{8}, &\mbox{(2D)}. 
\end{dcases}
\end{align}
The dielectric function for 3D and 2D Dirac materials are given in Refs.~\cite{Abrikosov70,Lv13,Johannes15b} and Refs.~\cite{Hwang07b,Wunsch06}, respectively. Using Eqs.~\eqref{eq:prop} and ~\eqref{eq:dielec}, we can then calculate the self-energy as
\begin{align}
\hat{\Sigma}_s(0,\mathbf{1}) &= 
\begin{dcases}
\frac{2s}{g} \ln\left(1+\frac{g \alpha}{3\pi} \ln \ell\right)  &\mbox{(3D)}\\
[f_0(\lambda) + s f_1(\lambda)] \ln \ell  &\mbox{(2D)}
\end{dcases}
+ \mathrm{finite},
\end{align}
where 
\begin{subequations}
\begin{align}
f_0(\lambda) &= -\frac{\alpha }{2\pi} \int_0^\pi d\chi \frac{\cos 2\chi}{1+\lambda \sin \chi} \nonumber\\
&= -\frac{\alpha}{\pi \lambda}
\left(-2 + \frac{\pi}{\lambda} - \frac{2-\lambda^2}{\lambda\sqrt{1-\lambda^2}}\arccos\lambda  \right),\\
\mbox{and}\nonumber\\
f_1(\lambda) &= \frac{\alpha}{2\pi} \int_0^\pi d\chi \frac{\cos^2\chi}{1+\lambda \sin \chi} \nonumber\\
&= \frac{\alpha}{\pi \lambda}\left(-1+\frac{\pi}{2\lambda}-\frac{\sqrt{1-\lambda^2}}{\lambda}\arccos\lambda \right).
\end{align}
\end{subequations}
The above derivation is given in Ref.~\cite{Johannes2}. The results obtained above are consistent with those given in Refs.~\cite{Son07,Gonzalez94,Gonzalez99} and Ref.~\cite{Gonzalez14} for 2D and 3D Dirac materials, respectively. 

\begin{figure}
\capstart
\begin{center}
\includegraphics[width=\linewidth]{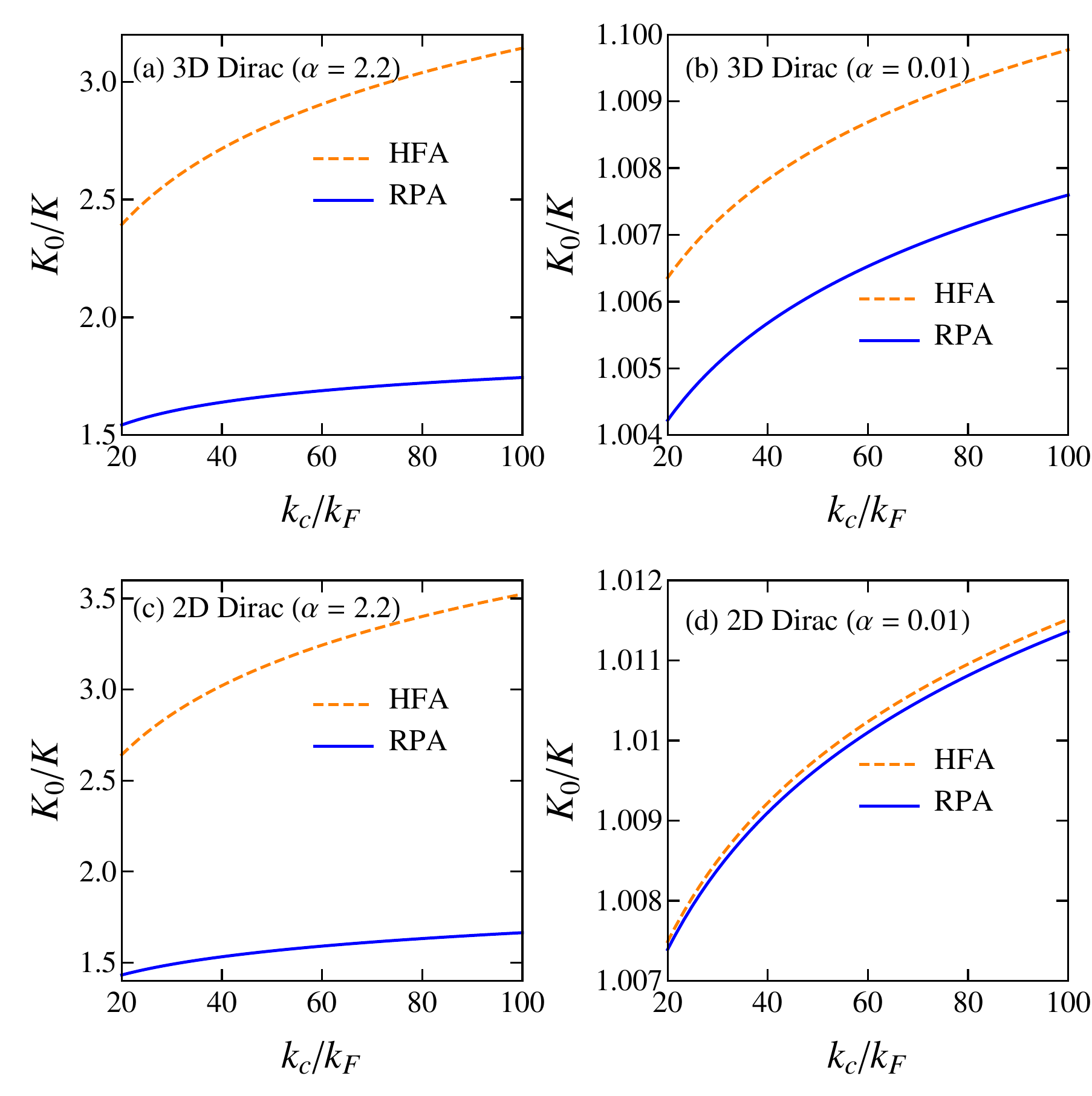}
\end{center}
\caption{(Color online) (Top) Zero-temperature incompressibility of 3D Dirac materials as a function of wave vector cutoff $k_c/k_F$ for different values of effective fine-structure constant: (a) $\alpha = 2.2$ and (b) $0.01$. (Bottom) Zero-temperature incompressibility of 2D Dirac materials as a function of wave vector cutoff $k_c/k_F$ for  (c) $\alpha = 2.2$ and (d) $0.01$.  The above plots are calculated using the degeneracy factor $g = 4$. }\label{fig:figRPA}
\end{figure}

\begin{figure}
\capstart
\begin{center}
\includegraphics[width=\linewidth]{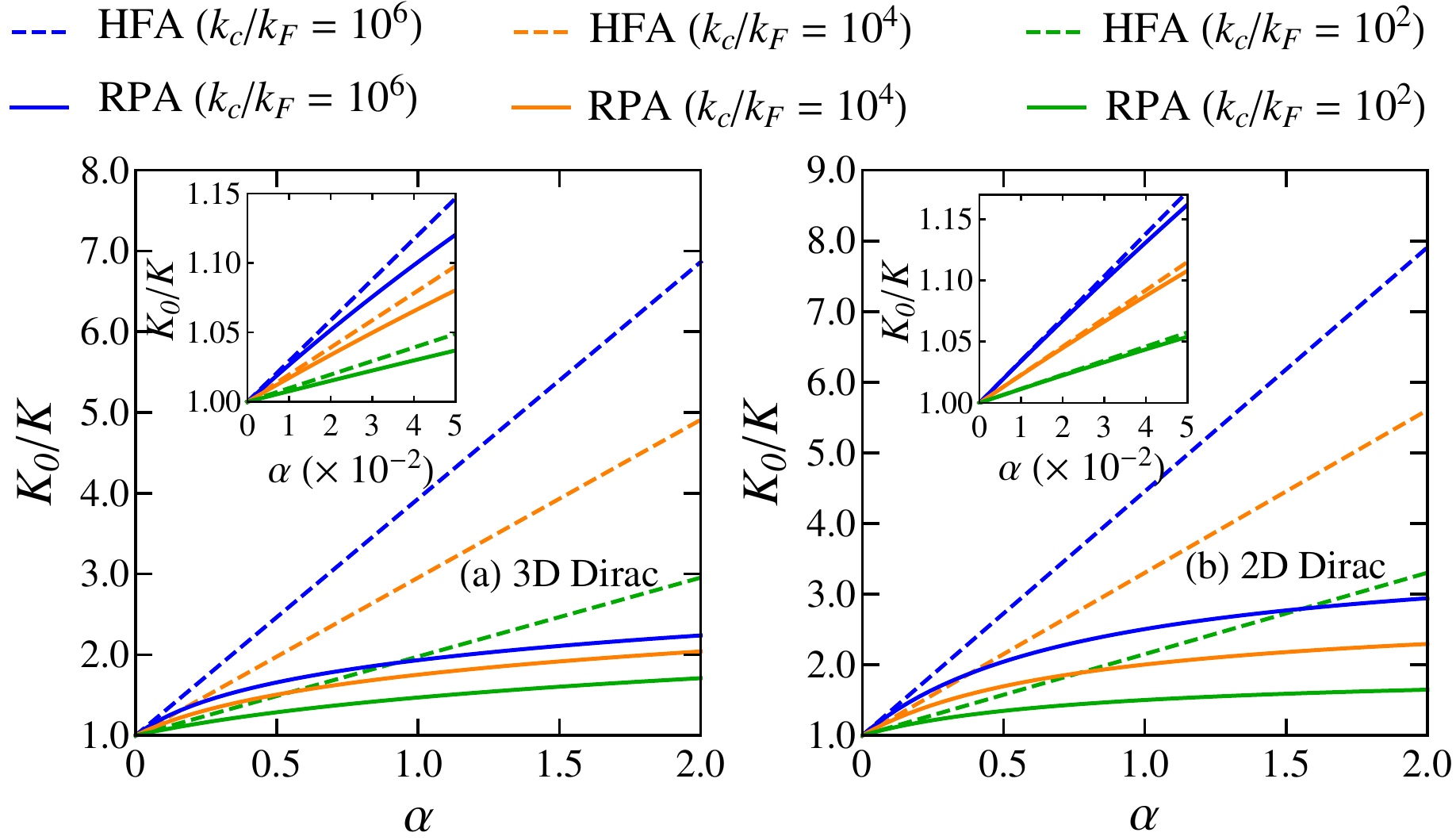}
\end{center}
\caption{(Color online) Zero-temperature incompressibility as a function of $\alpha$ for several different values of $k_c/k_F$ of (a) 3D Dirac and (b) 2D Dirac materials. Inset: Plot of zero-temperature incompressibility for small values of $\alpha$. The above plots are calculated using the degeneracy factor $g =4$. }\label{fig:RPAcomp}
\end{figure}

Considering extrinsic Dirac materials with $E_F > 0$ $(s = +1)$, we then obtain the zero-temperature RPA incompressibility (up to the divergent part of the interacting term) as
\begin{align}\label{eq:RPAcomp}
\frac{K_0}{K} =
\begin{dcases}
1 +  \dfrac{2}{g}\ln\left(1+\dfrac{g \alpha }{3\pi}\ln \dfrac{k_c}{k_F} \right) -\dfrac{2\alpha}{3 \pi} \dfrac{1}{1+\dfrac{g \alpha}{3\pi}\ln \dfrac{k_c}{k_F}} &\mbox{(3D)},\\
1 +  \left[f_2\left(\dfrac{g \pi \alpha}{8}\right)\right]\ln \dfrac{k_c}{k_F} &\mbox{(2D),}
\end{dcases}
\end{align}
 where $f_2(\lambda) = \frac{\alpha}{\pi \lambda}\left(1 -\frac{\pi}{2\lambda} + \frac{1}{\lambda \sqrt{1-\lambda^2}} \arccos \lambda\right)$. For the purpose of comparison, below we give the zero-temperature HF incompressibility (up to the leading-order divergent term in the interacting part) for 3D and 2D Dirac materials:
\begin{align}\label{eq:HFAcomp}
\frac{K_0}{K} = 
\begin{dcases}
1 + \dfrac{2\alpha}{3\pi} \ln\dfrac{k_c}{k_F} &\mbox{(3D)},\\
1 + \dfrac{\alpha}{4}\ln\dfrac{k_c}{k_F}  &\mbox{(2D).}
\end{dcases}
\end{align}

So within the RPA, the incompressibility $K_0/K$ for a 3D Dirac material increases slowly with $k_c/k_F$ with a $\log-\log$ dependence while $K_0/K$ for a 2D Dirac material increases as $\ln (k_c/k_F)$. Within the HFA, the incompressibilities $K_0/K$ for 2D and 3D Dirac materials increase as $\ln(k_c/k_F)$. The dependence of the zero-temperature HF and RPA incompressibilities on the momentum cutoff $k_c/k_F$ is shown in Fig.~\ref{fig:figRPA}. Figure~\ref{fig:RPAcomp} shows the plots of zero-temperature incompressibilities as functions of the effective fine-structure constant $\alpha$. As shown in the figure, the HF incompressibility increases linearly with $\alpha$ while the RPA compressibility increases with a rate that becomes smaller as $\alpha$ increases where the exact dependence given in Eq.~\eqref{eq:RPAcomp}. Note that for $\alpha \ll 1$, the leading-order divergent piece of the HF and RPA compressibilities are the same. This can be seen from the slope of the plots depicted in the right panel of Fig.~\ref{fig:figRPA} and insets of Fig.~\ref{fig:RPAcomp}. The fact that the leading-order interacting results in the coupling constant (for $\alpha \ll 1$) of the compressibility  agree for RPA and HFA indicates that we have no way of ascertaining which one is a better approximation within the single-loop theory, and experimental results are necessary in order to make progress.

\bibliographystyle{plain}

\end{document}